\documentclass[showpacs]{revtex4}
\usepackage{amssymb}

\newcommand{\beq}{\begin{eqnarray}}
\newcommand{\eeq}{\end{eqnarray}}
\newcommand{\nneeq}{\nonumber \end{eqnarray}}

\newcommand{\nn}{\nonumber \\}
\newcommand{\es}{& = &}
\newcommand{\rs}{\, = \,}
\newcommand{\ps}{& + &}

\newcommand{\np}{\nn \ps}

\newcommand{\cD}{ {\cal D} }
\newcommand{\cM}{ {\cal M} }
\newcommand{\cH}{ {\cal H} }

\newcommand{\cG}{ {\cal G} }
\newcommand{\cT}{ {\cal T} }

\newcommand{\cU}{ {\cal U} }
\newcommand{\cL}{ {\cal L} }
\newcommand{\cP}{ {\cal P} }

\begin{document}
\title{    Renormalization group procedure for effective particles:\\
           elementary example of exact solution with \\ finite 
           mass corrections and no involvement of vacuum 
           \footnote{Presented in part at the 50th Schladming 
           School of Theoretical Physics, Schladming, Austria, 25 February 
           - 3 March, 2012.}\footnote{Preprint number IFT/12/02} }
\author{   Stanis{\l}aw D. G{\l}azek }
\affiliation{  
           Institute of Theoretical Physics,
           Faculty of Physics, 
           University of Warsaw }
\date{     28 April, 2012}
\begin{abstract}
Renormalization group procedure for effective
particles in the front form of Hamiltonian
dynamics is applied to an elementary quantum 
field theory for two species of particles mixed 
through a mass-like interaction term. The model
interaction generates only finite terms and the 
procedure yields a whole family of equivalent 
effective theories. The exact solution for the 
family is found without involvement of the 
vacuum state in the dynamics. Physical spectrum 
is obtained at the end of the procedure in the 
form of free particles with definite masses. 
Since the procedure is designed in general terms, 
it could be used for the purpose of constructing 
effective dynamics also in other theories than 
the elementary model.
\end{abstract}
\maketitle

\section{ Introduction }

Renormalization group procedure for effective
particles (RGPEP)~\cite{npRGPEP} in the front form
(FF) of Hamiltonian dynamics~\cite{DiracFF} is
designed for application in solving relativistic
quantum field theories. In realistic theories,
where non-perturbative solutions of the RGPEP
equations are hard to find, the equations can be
initially solved only order-by-order in a
perturbative expansion~\cite{PnpRGPEP} or after
making other drastic simplifications of unknown
accuracy. This article describes instead an
application of the RGPEP to an elementary but
instructive model of a quantum field theory in
four dimensions which is soluble exactly. The
model exhibits a non-perturbative scale-evolution
of a mass matrix for effective particles and 
shows how the RGPEP can deal with the quantum
vacuum problem~\cite{DiracDeadWood}. The final 
result of the RGPEP in the model is a free theory 
of the particles whose masses appear in the
exact eigenvalues of the initial Hamiltonian. 

Since the RGPEP involves basic elements of the
canonical field quantization~\cite{QFT1,QFT2} and
renormalization of Hamiltonians by techniques
other than integrating out high-energy degrees of
freedom~\cite{Wilson1,Wilson2}, the elementary
model application is described including all
details needed to make the presentation
self-contained. Besides the FF of Hamiltonian
dynamics, the description often refers to the
commonly used form of dynamics, designated the 
instant form (IF) by Dirac~\cite{DiracFF}.

Section~\ref{RGPEP} briefly introduces the RGPEP.
The elementary example is defined in
Section~\ref{MH}. Section~\ref{SRGPEP} describes
solution of the RGPEP equations. The vacuum
problem is discussed in Section~\ref{vacuum}.
Section~\ref{conclusion} concludes the article.
Appendix \ref{massess} describes derivation 
of the same physical solution but obtained 
using an alternative RGPEP generator to the 
one used in the main text.

\section{ Summary of RGPEP }
\label{RGPEP}

The concept of effective particles as degrees of
freedom in a relativistic quantum field theory 
is introduced through a transformation~\cite{PnpRGPEP}
\beq
\label{qs}
\psi_s \es \cU_s \, \psi_0 \, \cU_s^\dagger \, .
\eeq
$\psi_s$ is a quantum field operator built from
creation and annihilation operators for effective 
particles of size $s$. These creation and annihilation 
operators are commonly denoted by $q_s$. The operator 
$\psi_0$ is the field operator built from the particle 
operators $q_0$ that correspond to the bare, point-like 
particles, and $s=0$. By definition, all kinematical 
quantum numbers that label operators $q$ on both sides 
of Eq.~(\ref{qs}), such as a three-momentum, charge,
spin, isospin, flavor, color, and the like, are 
not altered by $\cU_s$. 

The intuitive interpretation of parameter $s$ in
terms of a size of the effective particles in the
elementary model will be explained later. It is
based on the general RGPEP feature that effective
interactions contain the form factors that limit
how far off energy shell the interactions can
extend. The corresponding energy-width of the form
factors is determined by $1/s$ (see below). The
value $s=0$ corresponds to absence of form
factors. For a finite $s$, the effective
Hamiltonian is band-diagonal on the energy scale
and the band width is $\sim 1/s$. The principle of
using the band-diagonal structure for the purpose
of renormalization is formulated
in~\cite{GlazekWilson1}. It is convenient to use
the parameter $t=s^4$ and label operators with $t$
rather than $s$ itself.

A canonical Hamiltonian density is built from
products of fields $\psi_0$ and their derivatives.
A corresponding Hamiltonian is obtained by
integrating the density over a space-time
hyper-surface. The result is a polynomial
$\cH_0(q_0)$ with coefficients $c_0$. If a term in
$\cH_0(q_0)$ contains a product of $n$ operators
$q_0$, the coefficient has $n$ arguments. Each
argument is a set of quantum numbers carried by a
corresponding particle. Similarly, an
effective-particle Hamiltonian $\cH_t(q_t)$ is
defined through its coefficients $c_t$.
 
The RGPEP employs the equality
\beq
\cH_t(q_t) \es \cH_0(q_0) \, ,
\eeq
which means that the same dynamics is expressed in
terms of different operators. The change of $q_0$ to 
$q_t$ is accompanied with the change of coefficients 
$c_0$ to $c_t$ so that the physics is not changed. 
For example, the expansion of eigenstates of the
Hamiltonian into the $t$-dependent Fock components 
involves the wave functions that depend on $t$, but 
the states as a whole do not depend on $t$ at all. 

Variation of the coefficients $c_t$ with $t$
is described by the equation obtained by 
differentiating both sides of 
\beq
\cH_t(q_0) \es \cU^\dagger_t \, \cH_0(q_0) \, \cU_t \, ,
\eeq
with respect to $t$. One obtains
\beq 
\label{ht1}
\cH'_t(q_0) \es
[ \cG_t(q_0) , \cH_t(q_0) ] \, ,
\eeq 
where $\cG_t = - \cU_t^\dagger \cU'_t$ is called 
a generator. Correspondingly, 
\beq
\label{Usolution}
\cU_t 
\es 
T \exp{ \left( - \int_0^t d\tau \, \cG_\tau \right) } \, ,
\eeq
where $T$ orders operators from left to right 
in the order from a smallest to largest $t$. 

In the RGPEP, the generator is defined by
\beq
\label{cGdef}
\cG_t \es [ \cH_f, \cH_{Pt} ] \, .
\eeq
The operator $\cH_f$, called the free Hamiltonian, 
is the part of $\cH_0(q_0)$ that does not depend
on the coupling constants, 
\beq
\label{cHf} 
\cH_f \es
\sum_i \, p_i^- \, q^\dagger_{0i} q_{0i} \, .
\eeq 
The sum over subscript $i$ extends over all 
particle species and their quantum numbers, 
including integration over momenta, and 
\beq
p^-_i \es { p_i^{\perp \, 2} + m_i^2 \over p_i^+} \, .
\eeq
This is the FF free-energy of a particle with 
mass $m_i$ and kinematical momentum components 
$p_i^+$ and $p_i^\perp$. The operator $\cH_{Pt}$ 
is defined using the Hamiltonian $\cH_t$.
Namely, if $\cH_t(q_0)$ is of the form 
\beq
\label{Hstructure} 
\cH_t(q_0) \es
\sum_{n=2}^\infty \, 
\sum_{i_1, i_2, ..., i_n} \, c_t(i_1,...,i_n) \, \, q^\dagger_{0i_1}
\cdot \cdot \cdot q_{0i_n} \, ,
\eeq 
where the coefficients $c_t(i_1,...,i_n)$ are to 
be found using RGPEP, the operator $\cH_{Pt}(q_0)$ is 
defined by
\beq
\label{HPstructure} 
\cH_{Pt}(q_0) \es
\sum_{n=2}^\infty \, 
\sum_{i_1, i_2, ..., i_n} \, c_t(i_1,...,i_n) \, \,\left( {1 \over
2}\sum_{k=1}^n p_{i_k}^+ \right)^2 \, \, q^\dagger_{0i_1}
\cdot \cdot \cdot q_{0i_n} \, .
\eeq 
This means that ${\cal H}_{Pt}$ differs from ${\cal
H}_t$ by multiplication of each and every term by 
a square of a total $+$ momentum involved in a term.
In summary, the coefficients $c_t$ of products of
operators $q_t$ in the effective Hamiltonians 
$\cH_t(q_t)$, are solutions of the equation 
\beq 
\label{tnpRGPEP}
\cH'_t \es
\left[ [ \cH_f, \cH_{Pt} ], 
\cH_t \right] \, ,
\eeq 
where all operators are written as polynomials
in $q_0$ and the initial condition is provided
by a regulated canonical Hamiltonian with 
counterterms.

The counterterms are calculated in the RGPEP using
a condition that for finite $t$ the coefficients
$c_t$ with finite arguments do not depend on the
regularization parameters used in the canonical
Hamiltonian~\cite{GlazekWilson1}. The difficulty
of satisfying the cutoff-independence condition
for $c_t$ originates in the fact that the
coefficients appear in the solution for
$\cH_t(q_t)$ while the counterterms are inserted
in the initial condition $\cH_0(q_0)$ and
in-between there is a solution of the RGPEP that
spans the range from 0 to $t$. However, there is
no special difficulty associated here with the
counterterms because the coefficients $c_t$ with
finite arguments do not develop any dependence on
regularization in the model and solutions for them
are known exactly. Therefore, the adjustment of
counterterms in the model only amounts to
specifying their finite parts. These parts form
the initial mass matrix. The only regularization
dependence in the example appears in one overall
constant term in $\cH_t$, which is a pure number
and drops out from Eq.~(\ref{tnpRGPEP}).

The generic feature of narrowness of $\cH_t$ as $t$ 
increases can be seen by introducing a projector $R$ 
on a subspace in the Fock space. Let $\cH_R = R \, 
\cH_t R$. The corresponding projected equation reads 
(for details, see Appendix C in~\cite{npRGPEP}) 
\beq 
\label{narrowR2}
\cH_R' \es
\left[ [\cH_f, \cH_{PR}], \cH_R \right] \, .
\eeq 
The free Hamiltonian $\cH_f$ commutes with $R$.
The matrix version of Eq.~(\ref{narrowR2}) resembles 
the Wegner flow equation introduced in the IF of 
dynamics for Hamiltonians in condensed matter 
physics~\cite{Wegner1,Wegner2,Kehrein}. Similar 
equations are also successfully used in nuclear 
physics~\cite{OSU1,FurnstahlReview,FurnstahlLocal}. 
In relativistic quantum field theories, a narrow 
matrix must be obtained from Eq.~(\ref{narrowR2}) for
large $t$ because the trace of $\cH_R^2$ does not 
depend on $t$ and thus~\cite{npRGPEP} 
\beq
\label{start6x}
\left( \sum_{mn} |\cH_{Imn}|^2 \right)'
\es
- 2 \sum_{km} (\cM^2_{km} - \cM^2_{mk})^2
|\cH_{Ikm}|^2 \le 0 \, ,
\eeq
where $\cM_{km}$ denotes an invariant mass of
the particles in a state labeled with $k$ that 
are connected through the interaction $\cH_I$  
to the particles in a state labeled by $m$. The
interaction Hamiltonian is defined by $\cH_I 
= \cH - \cH_f$ and the matrix elements 
$\cH_{mn} = \langle m | \cH | n \rangle$ are
evaluated in the basis built from eigenstates 
$|m\rangle$ of $\cH_f$. Eq.~(\ref{start6x}) 
means that the sum of moduli squared of all 
matrix elements of the interaction Hamiltonian 
decreases as $t$ increases until all off-diagonal 
matrix elements of the interaction Hamiltonian 
between states with different free invariant 
masses vanish. For sizable value of $s$, the 
width of the narrow invariant-mass band in 
$\cH_R$ is $s^{-1}$. 

\section{ Model Hamiltonian }
\label{MH}

Let a theory of two real scalar fields $\phi$ 
and $\chi$ have a classical Lagrangian density
\beq
\label{lagrangian}
\cL \es {1 \over 2} \left[ (\partial \phi)^2 - \mu^2 \phi^2 \right] 
    +   {1 \over 2} \left[ (\partial \chi)^2 - \nu^2 \chi^2 \right] 
    -   m^2 \, \phi \, \chi \, .
\eeq
The last term is called the mass mixing term. 

\subsection{ Classical Hamiltonian }
\label{CH}

In terms of the variables $x^\pm = x^0 \pm x^3$ 
and $x^\perp = (x^1,x^2)$ used to label points 
in space-time, so that $\partial^\pm = 2\partial/
\partial x^\mp$, the Lagrangian density reads
\beq
\cL \es 
  {1 \over 2} \left[  \partial^+ \phi \, \partial^-\phi 
                   - (\partial^\perp \phi)^2  -   \mu^2 \phi^2 \right] 
+ {1 \over 2} \left[  \partial^+ \chi \, \partial^-\chi 
                   - (\partial^\perp \chi)^2  -   \nu^2 \chi^2 \right] 
    -   m^2 \phi \chi \, .
\eeq
The FF of dynamics involves the four-momentum~\cite{Yan1,Yan2} 
\beq
P^\mu        \es {1 \over 2} \int dx^- d^2 x^\perp \, \cT^{+\mu}(x) \, ,
\eeq
where the energy-momentum tensor density component 
relevant for constructing the model Hamiltonian is
\beq
\cT^{+-}(x) \es  
\partial^+ \phi \, \partial^-\phi + \partial^+ \chi \, \partial^-\chi 
- 2 \cL \, .
\eeq
Hence,  
\beq
\label{CP-}
P^-  \es {1 \over 2} \int dx^- d^2 x^\perp \,
\left[ (\partial^\perp \phi)^2 + \mu^2 \phi^2  
     + (\partial^\perp \chi)^2 + \nu^2 \chi^2 
     +  2 m^2 \phi \chi \right] \, .
\eeq

\subsection{ Quantization }
\label{Q}

Let the fields $\phi$ and $\chi$ at $x^+=0$ have 
the Fourier decompositions~\cite{NonlocalH}
\beq
\label{phi}
\phi(x^-,x^\perp)
\es
\int [p] \, a_p \, e^{-ipx} \, , \\
\label{chi}
\chi(x^-,x^\perp)
\es
\int [p] \  b_p \, e^{-ipx} \, ,
\eeq
where $[p]$ denotes the measure $d^+ p \, d^2p^\perp/[2|p^+| 
(2\pi)^3]$ of integration over momentum variables 
$p^+ = p^0 + p^3$ and $p^\perp = (p^1,p^2)$. In this 
notation, the integration over momentum variables 
extends from $-\infty$ to $+\infty$ for all three 
components of $p$ (a need for a cutoff on the range 
of $p$ is still ignored at this point). Quantum 
theory is obtained by imposing commutation relations
\beq
\label{cr}
{[}a_p, a_q{]} \es  {[}b_p, b_q{]} \rs
2p^+(2\pi)^3 \delta^3(p+q) \, .
\eeq 
The absence of $\dagger$ in the commutation relations
is intended, since it is the sign of $p^+$ that 
distinguishes the operators that create field quanta 
from operators that annihilate them. Such kinematical
distinction between creation and annihilation processes
is not available in the standard, IF approaches. The 
``annihilation'' operators with negative $p^+$ correspond 
to creation operators and one has
\beq
\label{amp}
a_{-p}     \es  a^\dagger_p \, , \\
\label{bmp}
b_{-p}     \es  b^\dagger_p \, .
\eeq
Note that these relations involve the change of sign of
$p^\perp$.

After quantization, the classical fields $\phi$
and $\chi$ are turned into operators that create
and annihilate quanta on the front hyper-plane, 
$\hat \phi$ and $\hat \chi$. The commutation 
relations of Eq.~(\ref{cr}) correspond to the 
spatial commutation relations
\beq
{[} \hat \phi(x) , \partial^+ \hat \phi(y) {]} 
\es
{[} \hat \chi(x) , \partial^+ \hat \chi(y) {]} 
\rs  i \delta^3(x-y) \, .
\eeq
The inverse relations are
\beq
\label{ap}
a_p            \es  |p^+| \int d^3x \, e^{ + i p
\, x} \, \hat \phi(x) \, , \\
b_p            \es  |p^+| \int d^3x \, e^{ + i p
\, x} \, \hat \chi(x) \, , 
\eeq
where $d^3x = dx^- d^2 x^\perp$ and integrals 
extend from $-\infty$ to $+\infty$ on the 
$x^+=0$ hyper-plane (the behavior of fields in 
spatial infinity remains unspecified at this 
point).

\subsection{ Quantum Hamiltonian }
\label{QH}

The quantum Hamiltonian is obtained from Eq.
(\ref{CP-}) by inserting operator versions 
of Eqs.~(\ref{phi}) and (\ref{chi}) for $\hat 
\phi$ and $\hat \chi$, respectively, and by
normal ordering,
\beq
\label{Hquantum}
\cP^-  \es {1 \over 2} \int dx^- d^2 x^\perp \,
: \left[ (\partial^\perp \hat \phi)^2 + \mu^2 \hat \phi^2  
       + (\partial^\perp \hat \chi)^2 + \nu^2 \hat \chi^2 
     +  2 m^2 \hat \phi \, \hat \chi \right] : \, .
\eeq
The normal ordering is defined using Feynman's
convention~\cite{Feynman} with the ordering
parameter set equal to $p^+$. In this convention, 
it is understood that operators $a_p$ are ordered 
in products according to the value of $p^+$ so 
that the greater $p^+$ the further to the right 
the operator. 

All terms in the Hamiltonian are bilinear in fields
and all of them contain one and the same integral
\beq
\label{integral}
\int dx^- d^2 x^\perp \,
\int [q \, p] \, e^{-iq\,x -ip\,x} 
\es
\int [q \, p] \, 2(2\pi)^2 \delta^3(q+p) \, .
\eeq
According to Eq.~(\ref{Hquantum}),
\beq
\label{Hquantum1}
\cP^-  \es {1 \over 2} \int [p] \,{1 \over |p^+| } \,
: \left[ (p^{\perp \, 2} + \mu^2) \, a_p \, a_{-p}
       + (p^{\perp \, 2} + \nu^2) \, b_p \, b_{-p}
      +                     2 m^2 \, a_p \, b_{-p}
\right] : \, . 
\eeq
The normal ordering produces the operator that 
properly counts the FF energy of field quanta, 
\beq
\label{Hquantum4}
\cP^-  \es \int [p] \, \theta(p^+)
  \left[ { p^{\perp \, 2} + \mu^2 \over p^+ } \, a^\dagger_p \, a_p 
       + { p^{\perp \, 2} + \nu^2 \over p^+ } \, b^\dagger_p \, b_p
       + {                    m^2 \over p^+ } \, (a^\dagger_p \, b_p + b^\dagger_p \, a_p) \right] \, .
\eeq
The last term describes the mixing of bare 
particles of type $a$ associated with field 
$\phi$ and of type $b$ associated with field 
$\chi$. From now on, the function $\theta(p^+)$ is 
included in the integration measure $[p]$.

The diverging number that is removed by the 
FF normal ordering in $\cP^-$, 
\beq
\label{Omega-}
\Omega^- 
\es 2 (2\pi)^3 \delta^3(0) \int [p] \, (p^{\perp
\, 2} + \mu^2/2 + \nu^2/2) \, ,
\eeq
involves factors $V_F = 2 (2\pi)^3 \delta^3(0)$
and $\rho_\Omega = \int [p] \, (p^{\perp \, 2} +
\mu^2/2 + \nu^2/2)$. Factor $V_F$ has an 
interpretation of a volume of the front that a
Hamiltonian density is integrated over. Factor
$\rho_\Omega$ is associated with a ground-state 
energy, cf. \cite{Stevenson1,Stevenson2,Hari,BartnikGlazek,
RozowskyThorn}. As a number, $\Omega^-$ does not 
contribute to the commutators in Eq.~(\ref{tnpRGPEP}) 
and it is not included in the RGPEP discussion in 
the next section. However, regarding application 
of the RGEPEP to more complex theories, one should
remember that the vacuum issue is not limited in
them to a constant such as $\Omega^-$,
cf.~\cite{KogutSusskind, Wilsonetal}.

\section{ Solution of the RGPEP equation }
\label{SRGPEP}

According to Sec.~\ref{RGPEP}, Eq.~(\ref{tnpRGPEP})
has the form, 
\beq 
\label{LFRGPEP}
{\cP^-_t}' \es
\left[ [ \cP^-_f, \cP^-_{Pt} ], 
\cP^-_t \right] \, ,
\eeq 
and should be solved using Eq.~(\ref{Hquantum4})
as the initial condition, 
\beq
\label{P0}
\cP^-_0(a_0,b_0)  \es \int [p] \,
  \left[ { p^{\perp \, 2} + \mu^2 \over p^+ } \, a^\dagger_{0p} \, a_{0p} 
       + { p^{\perp \, 2} + \nu^2 \over p^+ } \, b^\dagger_{0p} \, b_{0p}
       + {                    m^2 \over p^+ } \,(a^\dagger_{0p} \, b_{0p} 
                                               + b^\dagger_{0p} \, a_{0p}) \right] \, .
\eeq

\subsection{ Equations for coefficients $c_t$ }
\label{eqct}

On the basis of hindsight, the relevant operators 
can be written as
\beq
\label{cPt}
\cP^-_t(a_0,b_0)  
\es 
\int [p] \,
  \left[ A_{tp}  \, a^\dagger_{0p}  \, a_{0p} 
       + B_{tp}  \, b^\dagger_{0p}  \, b_{0p} 
       + C_{tp}  \, (a^\dagger_{0p}  \, b_{0p}  + b^\dagger_{0p}  \, a_{0p}  ) \right] \, , \\
\label{cPf}
\cP^-_f(a_0,b_0)   
\es  
\int [p] \,
  \left( A_{0p} \, a^\dagger_{0p}  \, a_{0p}  
       + B_{0p} \, b^\dagger_{0p}  \, b_{0p}  \right) \, , \\
\label{cPPt}
\cP^-_{Pt}(a_0,b_0)   
\es  
\int [p] \, p^{+2} \,
 \left[  A_{tp} \, a^\dagger_{0p}  \, a_{0p}  
       + B_{tp} \, b^\dagger_{0p}  \, b_{0p} 
       + C_{tp} \,(a^\dagger_{0p}  \, b_{0p}  + b^\dagger_{0p}  \, a_{0p} ) \right] \, ,
\eeq
where the coefficients generically denoted by $c_t$ in Sec. \ref{RGPEP} read
\beq
\label{At}
A_{tp}    \es { p^{\perp \, 2} + \mu_t^2 \over p^+ } \, , \\ 
\label{Bt}
B_{tp}    \es { p^{\perp \, 2} + \nu_t^2 \over p^+ } \, , \\
\label{Ct}
C_{tp}    \es {                    m_t^2 \over p^+ } \, .
\eeq
The initial conditions for these coefficients, denoted by 
$c_0$ in Sec. \ref{RGPEP}, are set by fixing the mass-squared 
parameters at $t=0$,
\beq
\mu_0 \es \mu \, , \\ 
\nu_0 \es \nu \, , \\
m_0   \es m \, ,
\eeq
with constants $\mu$, $\nu$, and $m$, taken from 
$\cP_0^-$ in~Eq. (\ref{P0}). These parameters 
include the finite parts of mass-squared counterterms 
as discussed in Section \ref{RGPEP}.

In this notation, the generator has the form
\beq
\label{GP}
[ \cP^-_f, \cP^-_{Pt} ] 
\es
\int [p] \,      \left( A_{0p} - B_{0p} \right) \, 
p^{+2} \, C_{tp} \, \left( a^\dagger_{0p} \, b_{0p} -  b^\dagger_{0p} \, a_{0p} \right) \, .
\eeq
Eq.~(\ref{LFRGPEP}) reads
\beq
{\cP^-_t}'(a_0,b_0) 
\es
\int [p] \,
  \left[ A'_{tp}  \, a^\dagger_{0p} \, a_{0p}
       + B'_{tp} \, b^\dagger_{0p} \, b_{0p}
       + C'_{tp} \, \left( a^\dagger_{0p} \, b_{0p}
                       + b^\dagger_{0p} \, a_{0p} \right) \right] \\
\es
\int [p] \,(- p^{+2}) \, (A_{0p} - B_{0p}) \, (A_{tp} - B_{tp}) \,  
                         C_{tp}\, \left( a^\dagger_{0p} \, b_{0p}
                                       + b^\dagger_{0p} \, a_{0p} \right) \\
\np
\int [p] \, 2 p^{+2} \,  (A_{0p} - B_{0p}) \, C_{tp}^2 \,
( a^\dagger_{0p} \, a_{0p} - b^\dagger_{0p} \, b_{0p} ) \, .
\eeq
By equating coefficients in front of the same
bare particle operators (or evaluating matrix 
elements between bare one-particle states of 
types $a$ and $b$), one obtains a set of equations 
for the coefficients $A_{tp}$, $B_{tp}$, and $C_{tp}$ 
in $\cP_t^-(a_0,b_0)$. Namely,
\beq
\label{Atprime}
A'_{tp} \es   2 p^{+2}  \, (A_{0p} - B_{0p})                        \, C_{tp}^2 \, , \\
\label{Btprime}
B'_{tp} \es - 2 p^{+2}  \, (A_{0p} - B_{0p})                        \, C_{tp}^2 \, , \\
\label{Ctprime}
C'_{tp} \es ( - p^{+2}) \, (A_{0p} - B_{0p}) \, (A_{tp}  - B_{tp} ) \, C_{tp}  \, .
\eeq
This set contains as many triplets of equations 
as there are different triplets of momentum labels 
$p$, which a priori is an infinite number when one 
does not regulate the field expansions into their 
Fourier components by imposing cutoffs on some 
discretized set of variables $p^+$ and $p^\perp$. 
However, it is clear that the modes with different 
values of $p$ are decoupled. They evolve in $t$ 
independently of each other. This simplification 
is a consequence of the bilinear nature of the 
initial Lagrangian. In addition, new generic 
simplifications occur thanks to the FF boost 
invariance of Eq. (\ref{tnpRGPEP}). 

\subsection{ Generic simplification due to boost invariance}
\label{boostinvariance}

In full detail, Eqs.~(\ref{Atprime}), (\ref{Btprime}), and~(\ref{Ctprime}),
read
\beq
\left( { p^{\perp \, 2} + \mu_t^2 \over p^+ } \right)'  
\es   2 p^{+2} \, \left( { p^{\perp \, 2} + \mu^2 \over p^+ }  
                       - { p^{\perp \, 2} + \nu^2 \over p^+ } \right) \, 
                  \left( {  m_t^2 \over p^+ } \right)^2 \, , \\
\left( { p^{\perp \, 2} + \nu_t^2 \over p^+ } \right)'    
\es - 2 p^{+2} \, \left( { p^{\perp \, 2} + \mu^2 \over p^+ }  
                       - { p^{\perp \, 2} + \nu^2 \over p^+ } \right) \, 
                  \left( {  m_t^2 \over p^+ } \right)^2 \, , \\
\left( {  m_t^2 \over p^+ } \right)'  
\es (- p^{+2}) \, \left( { p^{\perp \, 2} + \mu^2 \over p^+ }  
                       - { p^{\perp \, 2} + \nu^2 \over p^+ }  \right) 
                  \left( { p^{\perp \, 2} + \mu_t^2 \over p^+ }  
                       - { p^{\perp \, 2} + \nu_t^2 \over p^+ } \right) \, {  m_t^2 \over p^+ } \, .
\eeq
It is visible that the kinematical variables 
$p^+$ and $p^\perp$ drop out. This feature is 
special to the FF of dynamics. Thus, the a 
priori infinite set of different equations for 
infinitely many coefficients with different 
kinematical variables $p$, actually reduces to 
a single set of just 3 equations for 3 mass
parameters that are independent of $p$,
\beq
\label{mu}
\left( \mu_t^2 \right)'   
\es   2 \, \delta \mu^2 \, 
                  \left( m_t^2 \right)^2 \, , \\
\label{nu}
\left( \nu_t^2 \right)'    
\es - 2 \, \delta \mu^2 \, 
                  \left( m_t^2 \right)^2 \, , \\
\label{m}
\left(   m_t^2 \right)'
\es -      \delta \mu^2 \, 
           \left( \mu_t^2    - \nu_t^2    \right)
\, m_t^2 \, ,
\eeq
where 
\beq
\delta \mu^2 \es \mu^2 - \nu^2 \, .
\eeq
This set can be written as a differential 
matrix equation,
\beq
\label{masssquared}
\left[ \begin{array}{cc}
               \mu_t^2  &    m_t^2      \\
                 m_t^2  &  \nu_t^2
               \end{array} \right]'
\es
\left[
\left[  \left[ \begin{array}{cc}
               \mu^2  &  0            \\
               0      &  \nu^2
               \end{array} \right],
\left[ \begin{array}{cc}
               0      &   m_t^2       \\
               m_t^2  &   0
               \end{array} \right] \right],
\left[ \begin{array}{cc}
               \mu_t^2 &  m_t^2      \\
                m_t^2  &  \nu_t^2
               \end{array} \right]
\right] \, ,
\eeq
for a $2 \times 2$ matrix that will be 
called mass-squared matrix below. 

Note that Eq.~(\ref{masssquared}) would be a
Wegner-like equation if the first matrix on 
the right-hand side contained $\mu_t$ and 
$\nu_t$ instead of the initial mass parameters 
$\mu$ and $\nu$. Such change corresponds to 
inserting $\mu_t^2$ and $\nu_t^2$ in place of 
$\mu^2$ and $\nu^2$, respectively, in $\cP_f$ 
of Eq.~(\ref{cPf}). The resulting Wegner-like 
equation for the mass-squared matrix can be 
solved by proceeding in a way analogous to the 
one described below. This is shown in 
Appendix~\ref{massess}. The explicit solution 
described in next sections is for constant 
masses in $\cH_f$. 

\subsection{ Analytic solution for masses }
\label{analytic}

One can introduce a dimensionless variable 
\beq
u \es \delta \mu^4 \, t \, ,
\eeq
and, denoting differentiation with respect to $u$ 
with a prime, one obtains Eqs.~(\ref{mu}), 
(\ref{nu}), and (\ref{m}), in the form
\beq
\label{alpha}
\alpha'   
\es   2 \, \gamma^2 \, , \\
\label{beta}
\beta'    
\es - 2 \, \gamma^2 \, , \\
\label{gamma}
\gamma'
\es  -      \left( \alpha - \beta \right) \, \gamma \, ,
\eeq
where the dimensionless functions of $u$ are 
\beq
\alpha \es \mu_t^2/\delta \mu^2 \, , \\
\beta  \es \nu_t^2/\delta \mu^2 \, , \\
\gamma \es   m_t^2/\delta \mu^2 \, .
\eeq
If $\mu^2 = \nu^2$, so that $\delta \mu^2 = 0$, the
mass parameters do not evolve with $t$ irrespective 
of the initial value of mass-mixing parameter $m$. 
It is assumed from now on that $\mu^2 > \nu^2$, so 
that $\delta \mu^2 > 0$.

Regarding the mass degeneracy in the initial
theories, one should observe that in order to
trigger an RGPEP evolution towards a solution 
when initially $\mu=\nu$, one has to introduce 
an artificial splitting of masses in $\cH_f$. 
For example, such splitting is needed in the 
case of local theories with massless bare 
particles and chiral symmetry. Two other 
physically important cases in which the mass 
degeneracy and its minimal lifting may play 
important roles as far as an application of 
RGPEP is concerned, are neutrinos in electroweak 
interactions and $u$ and $d$ quarks in QCD.

Eqs.~(\ref{alpha}) and (\ref{beta}) imply that 
the sum $\alpha + \beta$ as a function of $u$ 
is a constant. This constant, denoted by 
$\cT = T/\delta \mu^2$, results from the 
constancy of a trace of the mass-squared 
matrix, $T = m_1^2 + m_2^2$, where $m_1^2$ 
and $m_2^2$ denote its eigenvalues. The 
remaining coupled set of equations reads 
\beq
\label{delta}
\delta'   
\es   4 \, \gamma^2 \, , \\
\label{gamma1}
\gamma'
\es  - \delta \, \gamma \, ,
\eeq
where $\delta = \alpha - \beta$. Multiplying 
the first of these two equations by $2\delta$
and the second by $2\gamma$, one arrives at 
\beq
\label{delta2}
{\delta^2}'   
\es   8 \, \delta \, \gamma^2 \, , \\
\label{gamma2}
{\gamma^2}'
\es  - 2 \delta \, \gamma^2 \, ,
\eeq
and concludes that  
\beq
\epsilon^2 \es \delta^2 + 4 \gamma^2
\eeq
does not depend on $u$. In fact, 
\beq
\label{epsilon}
\epsilon^2 \es \cT^2 - 4 \cD \, ,
\eeq
where $\cD = D/\delta \mu^4$ and $D$ is the 
determinant of the mass-squared matrix. 
Hence, $\epsilon^2 = (m_1^2 - m_2^2)^2 /
\delta\mu^4$. Using the constant $\epsilon$, 
one can eliminate $\gamma^2$ from 
Eq.~(\ref{delta2}) to obtain
\beq
\label{delta3}
\delta' \es   \epsilon^2 - \delta^2 \, ,
\eeq
which is an ordinary differential equation. 
Since the difference between eigenvalues of 
a hermitian $2 \times 2$ matrix is never smaller 
than the difference between its diagonal matrix 
elements, one always has $\delta' > 0$ except 
when $\delta = \epsilon$ and the mass-squared 
matrix is diagonalized. Without any loss of
generality one can assume $\epsilon > 0$. 

Integrations of Eqs.~(\ref{delta3}) and 
then~(\ref{gamma1}) produce solutions 
for the elements of mass-squared matrix
as functions of $t$,
\beq
\label{mut}
\mu_t^2 \es {1 \over 2} \, (\mu^2 + \nu^2) + { 1 \over 2} \, \delta \mu^2_t \, , \\
\label{nut}
\nu_t^2 \es {1 \over 2} \, (\mu^2 + \nu^2) - { 1 \over 2} \, \delta \mu^2_t \, , \\
\label{deltamut}
\delta \mu^2_t  \es     \delta \mu^2 \, 
                      { \cosh { x_t } + \epsilon \sinh { x_t }
                \over
                        \cosh { x_t } + \epsilon^{-1} \sinh { x_t } } \, , \\
\label{mt}
        m_t^2   \es m^2 \, { 1  \over
                        \cosh { x_t } + \epsilon^{-1} \sinh { x_t } } \, , 
\eeq
where  $x_t =  \delta \mu^2 \, \delta m^2 \, t$.
Note that $\epsilon = \sqrt{1 + (2 m^2/\delta \mu^2)^2}$. 
For $t \rightarrow \infty$, one obtains 
\beq
\mu_\infty^2 \es m_1^2 \, , \\ 
\nu_\infty^2 \es m_2^2 \, , \\
  m_\infty^2 \es 0     \, . 
\eeq
These results mean that the RGPEP eventually produces 
a Hamiltonian for the two new species of particles
of types 1 and 2 that are free, i.e., they no 
longer mix due to interactions, and their masses
squared are given by the eigenvalues $m_1^2$
and $m_2^2$ of the initial mass-squared matrix.

\subsection{ Effective particles }
\label{EP}

The result of RGPEP is a family of Hamiltonians
$P_t = \cP_t(a_t, b_t)$ for $t \ge 0$, which is 
obtained from $\cP_t(a_0,b_0)$ in Eq.~(\ref{cPt}) 
by replacement of $a_{0p}$ and $b_{0p}$ by $a_{tp}$ 
and $b_{tp}$, respectively. The effective particle 
operators are obtained from Eq.~(\ref{qs}). Namely,
\beq
a_{tp} \es \cU_t \, a_{0p} \, \cU^\dagger_t \, , \\
b_{tp} \es \cU_t \, b_{0p} \, \cU^\dagger_t \, ,
\eeq
where $\cU_t$ is given in Eq.~(\ref{Usolution})
as a solution of
\beq
\label{U'}
\cU'_t 
\es 
- \cU_t \, [ \cP^-_f, \cP^-_{Pt} ] \, .
\eeq
The generator, i.e., the commutator on the 
right-hand side of Eq.~(\ref{U'}), is given 
in Eq.~(\ref{GP}). Using results of the 
previous section, the generator can be written 
as
\beq
\label{GP1}
[ \cP^-_f, \cP^-_{Pt} ] 
\es
 \delta \mu^2 \, m_t^2 \, \int [p] \,
\left( a^\dagger_{0p} \, b_{0p} -  b^\dagger_{0p} \, a_{0p} \right) \, .
\eeq
Boost invariance of the RGPEP thus yields the 
generator that is a product of a function of 
$t$ times a constant operator. The $t$-ordered 
exponential in Eq.~(\ref{Usolution}) is 
\beq
\cU_t \es \exp{ (\varphi_t A )} \, , 
\eeq
where
\beq
\label{ctsolution}
\varphi_t \es - \delta \mu^2 \int_0^t m_\tau ^2 \, d\tau \\
    \es
\label{ctsolution2}
          \arctan{ \sqrt{ \epsilon + 1 \over \epsilon - 1} } 
        - \arctan{ e^{x_t} \sqrt{ \epsilon + 1 \over \epsilon - 1} } \, , \\
A   \es \int [k] \, 
\left( a^\dagger_{0k} \, b_{0k} -  b^\dagger_{0k} \, a_{0k} \right) \, .
\eeq
The effective particle operators $a_{tp}$ 
and $b_{tp}$ are obtained from the formula
\beq
q_{tp} 
    \es
e^{   \varphi_t A } 
\, q_{0p} \,
e^{ - \varphi_t A } 
\eeq
with $q=a$ and $q=b$, respectively, using 
\beq
{[}A, a_{0p}{]} 
\es 
- b_{0p} \, , \\
{[} A, b_{0p}{]} 
\es
a_{0p} \, .
\eeq
Suppose there exists a combination
\beq
q_{0p} \es a_{0p} + z_1 \, b_{0p} \, , 
\eeq 
for which one has
\beq
[A, q_{0p}] \es z_2 \, q_{0p} \, ,
\eeq
where $z_1$ and $z_2$ are some complex numbers.
This is possible when $z_1 = z_2 = \pm \,i$ and 
\beq
e^{   \varphi_t A } 
\, q_{0p\pm} \,
e^{ - \varphi_t A } 
\es
e^{ \pm \,i \, \varphi_t} \, q_{0p \pm} \, .
\eeq
where $q_{0p\pm} = a_{0p} \pm i \, b_{0p}$.
Knowing that 
\beq
a_{0p} \es { 1 \over2} (q_{0p+} + q_{0p-}) \, , \\
b_{0p} \es {-i \over2} (q_{0p+} - q_{0p-}) \, , 
\eeq
one obtains
\beq
a_{tp} 
       \es  \cos{\varphi_t} \, a_{0p} - \sin{\varphi_t} \, b_{0p} 
\label{atp} \, , \\
b_{tp} 
       \es  \sin{\varphi_t} \, a_{0p} + \cos{\varphi_t} \, b_{0p}  
\label{btp} \, ,
\eeq
and the inverse relations
\beq
a_{0p} \es   \cos{\varphi_t} \, a_{tp} + \sin{\varphi_t} \, b_{tp}  \, , \\
b_{0p} \es - \sin{\varphi_t} \, a_{tp} + \cos{\varphi_t} \, b_{tp}  \, .
\eeq
Eqs.~(\ref{atp}) and (\ref{btp}) provide explicit 
definitions of annihilation operators for
effective particles corresponding to the RGPEP
parameter $t = s^4$. The corresponding relations 
for creation operators are obtained by hermitian 
conjugation. 

\subsection{ Interpretation of $s$ as the effective particle size }
\label{size}

The interpretation of parameter $s$ as a size of
effective particles requires explanation in the
context of our mass-mixing model because the
mass-mixing interaction does not change any
three-momentum that could be an argument of a form
factor whose width might be related to a concept
of a particle size. However, in more advanced
theories, interactions change an invariant mass of
the interacting particles when their relative
momenta change. To be specific, consider a fermion
of mass $m_f$ that emits a boson of mass $m_b$.
The associated change of invariant mass squared is
\beq
\cM^2_{fb,f} \es
\left( \sqrt{ m_f^2 + k^2} + \sqrt{ m_b^2 + k^2} \right)^2 
- m_f^2 \, .
\eeq
The exponential factor of the type $\exp{(- s^4 
\cM^4_{fb,f})}$ becomes $\exp{[- (2sk)^4]}$ for 
large $k$. This is the origin of interpreting the 
parameter $s$ as a size of effective particles in 
complex theories. Namely, only particles with small 
size $s$ can interact producing a large momentum 
$k$, cf. \cite{NonlocalH}. 

In the mass-mixing model, there is no change of
relative three-momentum involved. Instead, the
interaction strength $m_t^2$ in Eq.~(\ref{mt}) is
limited in strength roughly by $\exp{( - \delta
\mu^2 \delta m^2 s^4)}$. The change of interaction
strength comes solely from the change of a
particle mass. Therefore, the role of the
effective particle size parameter $s$ is reduced
to taming changes in the mass. The point-like,
bare particles at $s=0$ can change mass through a
mass-mixing interaction by arbitrary amounts that
are introduced in the initial $\cP^-$. But the
effective particles of large $s$ can change mass
only by amounts not exceeding $1/s$, as if the
motion of their constituents could not involve a
large excitation without breaking them apart.
Thus, when $s$ is large, the effective particles
can only change their masses by small amounts.
Eventually, when $s \rightarrow \infty$, they
cannot change mass at all, which means that they 
do not interact through a mass mixing term at all 
(see Section~\ref{spectrum} below). In any case, 
the RGPEP suggests that mass mixing in low-energy
effective theories should be small. Realistic
effective theories appear to share this feature.

\subsection{ Constance of the Hamiltonian }
\label{constance}

The effective Hamiltonian, $\cP^-_t = \cP^-_t(a_t,b_t)$, 
is obtained from $\cP^-_t(a_0,b_0)$ in Eq.~(\ref{cPt}) 
by replacing $a_{0p}$ and $b_{0p}$ in the latter by 
$a_{tp}$ and $b_{tp}$. The result is
\beq
\cP^-_t 
\es 
\int [p] \,
  \left[ A_{tp}  \, a^\dagger_{tp}  \, a_{tp} 
       + B_{tp}  \, b^\dagger_{tp}  \, b_{tp} 
       + C_{tp}  \, (a^\dagger_{tp}  \, b_{tp}  + b^\dagger_{tp}  \, a_{tp}  ) \right] \, , 
\label{cPtsolution}
\eeq
where the coefficients $A_{tp}$, $B_{tp}$, and $C_{tp}$
are given in Eqs.~(\ref{At}), (\ref{Bt}), (\ref{Ct}),
respectively, and the mass parameters in them are 
given in Eqs.~(\ref{mut}), (\ref{nut}), and (\ref{mt}).
Using Eqs.~(\ref{atp}) and (\ref{btp}), one
obtains
\beq
\cP^-_t 
\es \cP^-_0
+
\int [p] \,
  \left[ {\Delta \mu^2 \over p^+} \, a^\dagger_{0p}  \, a_{0p} 
       + {\Delta \nu^2 \over p^+} \, b^\dagger_{0p}  \, b_{0p} 
       + {\Delta   m^2 \over p^+} \, (a^\dagger_{0p}  \, b_{0p}  + b^\dagger_{0p}  \, a_{0p}  ) \right] \, , 
\eeq
where
\beq
\Delta \mu^2 \es \mu_t^2 c^2 + \nu_t^2 s^2 + 2 \, m_t^2 c s - \mu^2 \, , \\
\Delta \nu^2 \es \mu_t^2 s^2 + \nu_t^2 c^2 - 2 \, m_t^2 c s - \nu^2 \, , \\
\Delta   m^2 \es -(\mu_t^2 - \nu_t^2) cs + m_t^2 (c^2 - s^2) -  m^2 \, ,
\eeq
$s = \sin \varphi_t$, and $c = \cos \varphi_t$. Direct
inspection demonstrates that $\Delta \mu^2 =
\Delta \nu^2 = \Delta m^2 = 0$ for all values of
$t$ in the range from 0 to $\infty$, which means
that the operators $\cP^-_t = \cP^-_t(a_t,b_t)$
and $\cP^-_0 = \cP^-_0(a_0,b_0)$ are the same for
all values of $t$.

\subsection{ Spectrum of the theory }
\label{spectrum}

The initial Hamiltonian, $\cP^-_0$ in
Eq.~(\ref{P0}), is transformed as a result of the
RGPEP to $\cP_t^-$ in Eq. (\ref{cPtsolution}). At
the same time, the RGPEP secures equality $\cP^-_t
= \cP^-_0$, as shown in Section \ref{constance}.
Since the eigenvalues and eigenstates of $\cP^-_0$
and $\cP^-_t$ are identical, one can derive them
using any value of $t$ one wishes. The simplest to
discuss is the case of $t \rightarrow \infty$,
because in this case there is no mass mixing,
$m_\infty = 0$. The mixing vanishes in the limit 
$t \rightarrow \infty$ provided that initially 
$\mu \neq \nu$. This is assumed in what follows. 
The case of $\mu = \nu$ is addressed near the end 
of this section.

The effective theory with $t= \infty$ is a free
theory, with a correspondingly simple spectrum.
Details of the spectrum are described below for
two reasons. One reason is the completeness of the
article. The other reason is a preparation for the
discussion in Section \ref{vacuum} concerning the
ground state, or vacuum. Simplicity of the RGPEP
illustrated here is contrasted with complexity of
other approaches there.

In the limit of $t \rightarrow \infty$,  
\beq
\label{Pinfty}
\cP^-_\infty(a_\infty,b_\infty)  \es \int [p] \,
  \left[ { p^{\perp \, 2} + \mu_\infty^2 \over p^+ } 
         \, a^\dagger_{\infty p} \, a_{\infty p} 
       + { p^{\perp \, 2} + \nu_\infty^2 \over p^+ } 
         \, b^\dagger_{\infty p} \, b_{\infty p} \right] \, ,
\eeq
where
\beq
a_{\infty p} 
\es  \cos{\varphi_\infty} \, a_{0p} - \sin{\varphi_\infty} \, b_{0p} 
\label{ainftyp} \, , \\
b_{\infty p} 
\es  \sin{\varphi_\infty} \, a_{0p} + \cos{\varphi_\infty} \, b_{0p}  
\label{binftyp} \, ,
\eeq
and the angle $\varphi_\infty$ is 
\beq
\label{cinfty}
\varphi_\infty 
\es - \arctan{ \sqrt{ \epsilon - 1 \over \epsilon
+ 1} } \, .
\eeq
Note that this angle is the same as the
one in Eq.~(\ref{cinftyA}) that results 
from solving RGPEP equations with a 
different generator in Appendix~\ref{massess}.

Eigenvalues of the Hamiltonian in Eq.
(\ref{Pinfty}) are free energies of 
$n_{\infty 1}$ particles of mass $m_1$ 
and $n_{\infty 2}$ particles of mass $m_2$, 
each with some momentum components $p^+$ 
and $p^\perp$,
\beq
P^-_{\{p_{1i}, i=1,..., n_{\infty 1}\},
     \{p_{2j}, j=1,..., n_{\infty 2}\}}
\es
\sum_{i=1}^{n_{\infty 1}} {p_{1i}^{\perp \, 2} + m_1^2 \over p_{1i}^+} 
+
\sum_{j=1}^{n_{\infty 2}} {p_{2j}^{\perp \, 2} + m_2^2 \over p_{2j}^+} \, .
\eeq
The spectrum is degenerate. The eigenstates can 
be closely identified because the RGPEP provides 
expressions for the operators $a_\infty$ and 
$b_\infty$. A complete set of eigenstates (not 
normalized) is defined by writing 
\beq
\label{eigenstates}
|\{p_{1i}, i=1,..., n_{\infty 1}\},
 \{p_{2j}, j=1,..., n_{\infty 2}\}\rangle
\es
\prod_{i=1}^{n_{\infty 1}} a^\dagger_{\infty p_{1i}}
\prod_{j=1}^{n_{\infty 2}} b^\dagger_{\infty p_{2j}} |0\rangle \, ,
\eeq
where $|0\rangle$ denotes the vacuum state. The 
vacuum state is annihilated by all annihilation 
operators of all particles for all values of $t$
and one can treat $|0\rangle$ as one and the same 
state for all values of the parameter $t=s^4$.

Since the creation operators $a^\dagger_{\infty p}$ 
and $b^\dagger_{\infty p}$ are given by linear 
combinations of $a^\dagger_{0 p}$ and $b^\dagger_{0 p}$
implied by Eqs.~(\ref{ainftyp}) and (\ref{binftyp}) 
through hermitian conjugation, the eigenstates 
defined in Eq.~(\ref{eigenstates}) can also be
written as combinations of states created from the 
same vacuum state by products of the operators 
$a^\dagger_{0p}$ and $b^\dagger_{0p}$ with the  
corresponding momenta. The total number of particles 
in every resulting component of an eigenstate is 
the same. However, an eigenstate with definite
numbers $n_{\infty 1}$ and $n_{\infty 2}$ of 
effective particles with $t=\infty$ corresponds 
to a combination of states with varying numbers  
of initial particles, $n_{0 1}$ and $n_{0 2}$, 
that satisfy the condition $n_{0 1}+n_{0 2} = 
n_{\infty 1} + n_{\infty 2}$. If the total number 
of particles is large, a simple state of effective 
particles with $t=\infty$ is a complex mixture of 
many states made of bare particles corresponding 
to $t=0$. 

When $\mu = \nu$, the RGPEP does not change the
particle operators, since the generator is zero.
On the other hand, it is clear that a non-zero
mixing term $m^2$ causes the eigenvectors of
mass-squared matrix in a classical Lagrangian to
be definite combinations of the initial basis
vectors. In the quantum theory, in order to
generate a solution using the RGPEP, one may
introduce a small artificial difference between
the initial masses. When the initial mass
degeneracy corresponds to symmetry, the small
artificial difference that breaks the degeneracy
breaks also the symmetry. The RGPEP can be said to
use consequences of such small breaking to finesse
quantum symmetry-breaking solutions.

In summary, the RGPEP produces the spectrum in 
a simple way. However, the simplicity is to some 
extent deceptive because the RGPEP allows one to 
ignore questions concerning the vacuum state
$|0\rangle$. The next section discusses this
issue.

\section{ The vacuum problem }
\label{vacuum}

The vacuum problem appears in the quantization of
fields~\cite{QFT1,QFT2,DiracDeadWood}. One starts
with quantizing a free classical theory. This
renders a quantum theory of non-interacting
particles in terms of a free Hamiltonian $H_0$.
Interaction terms are added to $H_0$ in the form
of $H_I$. The latter can be constructed by
starting from local products of classical fields
multiplied by coupling constants and replacing the
classical fields with the quantized ones. The
vacuum problem becomes apparent when one attempts
to solve the eigenvalue problem for $H = H_0 +
H_I$. The problem is that $H_I$ takes eigenstates
of $H_0$ out of the Hilbert space. In particular,
the ground state of the free theory, denoted by
$|0\rangle$, is changed by $H_I$ to a state with
an infinite norm. The situation is further
discussed below using the mass-mixing example, in
which the vacuum problem appears in a similar way
as in the model used by Dirac to discuss the vacuum
problem~\cite{DiracDeadWood}.

\subsection{ Vacuum problem due to mass-mixing }
\label{vIF}

The parameter $m^2$ in the mixing term in the Lagrangian 
of Eq.~(\ref{lagrangian}) is treated as a coupling
constant. Setting $m=0$, one obtains a Lagrangian 
density of a free theory, 
\beq
\label{lagrangian0}
\cL_0
\es {1 \over 2} \left[ (\partial \phi)^2 - \mu^2 \phi^2 \right] 
+   {1 \over 2} \left[ (\partial \chi)^2 - \nu^2 \phi^2 \right] \, .
\eeq
The IF quantization of a free theory is well-known
and nothing new is said here about it except for
stressing one aspect that concerns the vacuum.
Namely, when one evaluates $H_0 = \int d^3x \,
\cH_0$, where the Hamiltonian density $\cH_0$ is
canonically obtained from $\cL_0$, the terms that
involve products of two creation or two
annihilation operators all cancel out, as desired.
This happens because of the free energy formulae,
$E_a^2(p) = \mu^2 + p^2$ and $E_b^2(p) = 
\nu^2 + p^2$, that are used in defining the
time derivatives, or canonical momenta for the
field variables. These energy formulae produce the
desired cancellations in the sum of terms
involving $\pi_\phi^2$, $\vec \nabla \phi^2$, and
$\mu^2 \phi^2$, and similarly for $\pi_\chi^2$,
$\vec \nabla \chi^2$, and $\nu^2 \chi^2$. The
resulting $H_0$ in the IF of dynamics has the form
\beq
\label{H0}
H_0(a_0,b_0)  \es \int [p]_a \, \sqrt{\mu^2 + p^2} \, a^\dagger_{0p} \, a_{0p} 
              +   \int [p]_b \, \sqrt{\nu^2 + p^2} \, b^\dagger_{0p} \, b_{0p}  \, .
\eeq
The integration measures obtain the subscripts $a$
and $b$ because of the energies in their denominators. 
The non-zero commutation relations are $[a_{0p}, 
a^\dagger_{0q}] = 2 E_a(p) \, (2\pi)^3 
\delta^3(\vec p - \vec q\,)$ and $[b_{0p}, 
b^\dagger_{0q}] = 2 E_b(p) \, (2\pi)^3 
\delta^3(\vec p - \vec q\,)$. Possible additional 
quantum numbers can be ignored here. 

An infinite constant $\Omega^0$ has been removed
by the IF normal ordering, analogous to the
constant $\Omega^-$ removed from the FF
Hamiltonian $\cP^-$, see Eqs.~(\ref{Hquantum4})
and (\ref{Omega-}). The constant $\Omega^0$ can be
subtracted this way~\cite{DiracDeadWood,
BjorkenDrell}, or it can also be included in
variational estimates of the ground-state energy
when interaction terms are taken into
account~\cite{Stevenson1,Stevenson2}. 

In a theory set up this way, the IF vacuum problem 
emerges in the model due to the interaction term,
\beq
\label{HI}
H_I
\es \int d^3x \, m^2 \, \phi \, \chi \\
\label{HI1}
\es \int { d^3 p  \over (2\pi)^3 } \, 
         { m^2 \over 4 E_a E_b }  
     \left( a^\dagger_{0p} \, b^\dagger_{0-p}
          + a^\dagger_{0p} \,         b_{0p}
          + b^\dagger_{0p} \,         a_{0p}
          +         a_{0p} \,         b_{0-p}
\right) \, .
\eeq
The result of action of $H_I$ on the vacuum state is
\beq
\label{HIdiv}
H_I |0\rangle 
\es \int { d^3 p  \over (2\pi)^3 } \, 
         { m^2 \over 4 E_a E_b }   
\, a^\dagger_{0p} \, b^\dagger_{0-p} \, |0\rangle \, .
\eeq

This state has an infinite norm. The infinity
occurs through two factors. One factor is the
volume of space in which the states with definite
three-momentum are normalized. In the case of a
state in Eq.~(\ref{HIdiv}), the three-momentum is
zero. Another source of infinity is the integral 
over all momentum labels $p$. This divergence 
results from the infinite number of momentum 
scales in the theory. 

Acting on the state in Eq. (\ref{HIdiv}) with 
$H_I$ again also generates infinity. Multiple 
action of $H_I$ creates further infinities. For 
example, the infinities appear in action of the 
evolution operator $U = \exp{(-i H t)}$ on 
$|0\rangle$, since $U$ involves all powers of 
$H_I$~\cite{DiracDeadWood}. 

If states of the theory are built starting from
$|0\rangle$, the mixing operator $H_I$ creates
infinities in all of them. Removal of the
infinites requires a cutoff on the range of
momentum $p$ in the Fourier expansions of fields
$\phi$ and $\chi$. However, every cutoff on the
momentum range in a theory violates the Lorentz
symmetry~\cite{DiracDeadWood}. One has to
re-design the quantization procedure in the
example in order to recover the quantum theory
that was straightforwardly found as a solution
using the RGPEP in previous sections. 

On the one hand, it is known in the elementary
model what needs to be done to solve it. On the
other hand, one can look at the model as sharing
some basic features with theories in which a more
complex $H_I$ is added to $H_0$ and it is not
known how to deal with the vacuum problem in them
beyond perturbation theory. Therefore, the model
is of interest as a potential source of ideas
about how to use the RGPEP to try to work around
the vacuum problem in complex theories and attempt
to break through the barriers that this problem
poses in general.

\subsection{ General scope of vacuum problems }
\label{GS}

Dirac pointed out that problems with vacuum may
require a re-interpretation of quantum field
theory~\cite{DiracDeadWood}. He argued for such
re-interpretation in the case of QED. Similar
divergences occur in the vacuum problem of QCD but
they cannot be as easily worked around as Dirac
suggested for QED~\cite{Wilsonetal}. 

One reason is that the coupling constant in QCD is
much larger than in QED. The QED coupling constant
is so small that one can use very large cutoffs in
diverging terms in perturbation theory and still
does not need to worry about the Lorentz-symmetry
violation in practice. In QCD, where the coupling
constant is much larger than in QED, the cutoffs
would have to be much smaller than in QED in order
to exclude large terms in perturbation theory. But
much smaller cutoffs on $|\vec p\,|$ could lead to
effects that violate the Lorentz symmetry much
stronger. Asymptotic freedom enables perturbative
calculations in QCD but does not solve the vacuum
problem. The other reason is the need for
explaining spontaneous chiral symmetry
breaking~\cite{Nambu,GasserLeutwyler} for which a
non-trivial vacuum structure is seen as the
origin. The third reason is the desire to explain
confinement. Confinement is often associated in
the literature with a concept of a complex ground
state. In any case, the ground state of QCD still
awaits a construction. More generally, questions
concerning a ground-state, spontaneous symmetry
breaking, and mass generation, are of concern in
the present standard model and theories trying to
explain its origin. A famous ambiguity involved in
the vacuum concept is the vacuum energy density,
which can be seen as relevant to
cosmology~\cite{WeinbergCosmology1,
WeinbergCosmology2}. 

In the FF of Hamiltonian dynamics, the vacuum
problem does not appear in the same way as in 
the IF. For example, the vacuum problem in
the FF version of QCD can be formulated as a
renormalization group problem for Hamiltonians
\cite{Wilsonetal}. Using the RGPEP, one can also
envision a scenario for solving the canonical FF
of QCD in which the effects commonly associated
with a gluon condensate in vacuum~\cite{SVZ} may
actually originate in an analogous expectation
value but merely in the gluon medium that exists 
only inside the volume of a hadron, rather than 
in the entire space~\cite{npRGPEP}. Discussions 
of the idea that condensate parameters may actually 
correspond to expectation values of operators in 
the medium present inside hadrons, instead of the 
entire space, are available in~\cite{Marisqq,
Brodskyqq,BrodskyShrock,BrodskyRobertsSchrockTandy}, 
including implications for cosmology. 

The scope of vacuum problems is broad enough to
suggest that the features that enable RGPEP to
work around the vacuum problem and produce an 
exact quantum solution in the elementary example, 
should be identified. This is done in the next 
section.

\subsection{ RGPEP path around the vacuum }
\label{aroundv}

The general features that enable RGPEP to circumvent 
the vacuum problem and still produce a relativistic 
solution in the model stem from the FF of Hamiltonian 
dynamics. The key properties of the FF are the positivity 
of $p^+$ and boost invariance. The RGPEP takes advantage 
of these properties in the design of its generator.

The positivity of $p^+$ results from the assumption 
that for a free particle of an arbitrary mass 
$\mu > 0$ one can write for arbitrary three-momentum 
$\vec p$ that
\beq
\label{p+}
p^+ \es \sqrt{ \mu^2 + \vec p^{\,2} } + p^z \geq 0 \, .
\eeq
Thus, one assumes in the FF of quantum dynamics that 
a creation operator for a particle may only carry 
positive $p^+$ as a label. This feature is summarized 
in Eqs.~(\ref{amp}) and (\ref{bmp}) in Section \ref{Q}. 

Positivity of $p^+$ in Eq. (\ref{p+}) implies that 
the classical, translation-invariant mass-mixing 
interaction term in Eq.~(\ref{CP-}),
\beq
P^-_I \es \int dx^- d^2 x^\perp \, m^2 \phi \, \chi \, ,
\eeq
results in the quantum interaction operator in Eq.~(\ref{P0}),
\beq
\label{cP-I}
\cP^-_I  \es \int [p] \, \theta(p^+)
         { m^2 \over p^+ } \, (a^\dagger_{0p} \,
b_{0p} + b^\dagger_{0p} \, a_{0p}) \, ,
\eeq
which does not contain any terms of the type
$a^\dagger_p \, b^\dagger_{-p}$ and $a_p \,
b_{-p}$ that appear in Eq.~(\ref{HI1}) for 
$H_I$ in the IF of quantum dynamics. 
Such terms are excluded because both $p^+$ and 
$-p^+$ in them are required to be positive. This 
is not possible for particles of a finite mass 
in a presence of a cutoff on $p = |\vec p\,|$, 
as is visible in Eq.~(\ref{p+}), no
matter how large such cutoff is. Note also that
the FF integration measure $[p]$ does not depend
on the mass $\mu$ used in the condition
(\ref{p+}).
 
When the cutoff on $|\vec p\,|$ can be made
arbitrarily large, one can have boost invariance
in practice in an arbitrarily large range of
momenta provided that the theory respects the
symmetry~\cite{Wilsonetal}. This is the case at
$s=0$ in the RGPEP. In order to maintain the
Lorentz symmetry in an effective theory, the
sliding cutoff parameter $\lambda = 1/s$ emerges
in the RGPEP through its equations. They are 
so designed that the sliding cutoff is not 
limiting $|\vec p\,|$ of individual particles. 
Instead, the effective-theory cutoff limits 
only the changes of invariant mass caused by
interactions. The mass is invariant with 
respect to all 7 FF kinematical symmetries,
including boost invariance.

The boost invariance is secured by design of the
RGPEP generator in Eq.~(\ref{cGdef}). The
commutator guarantees that only connected
interactions are generated. The total transverse
momenta of interacting particles before and after
an interaction cancel each other in the arguments
of resulting vertex form factors. Spectators do
not contribute to these arguments. The
multiplication by a total $+$ momentum squared of
interacting particles in the definition of
$\cH_{Pt}$, Eq.~(\ref{HPstructure}), results in
the factor $p^{+ \,2}$ in Eq.~(\ref{cPPt}). In the
absence of sensitivity to cutoffs on $p^+$, this
factor removes $p^+$ from the RGPEP evolution
equation entirely. Therefore, the arguments of
resulting vertex form factors depend only on the
change of invariant mass squared among the
particles that are involved in the interaction. 

These features, combined with the absence of 
divergences due to separation of momentum modes, 
reduce the RGPEP in the elementary mass-mixing 
model to solving an evolution equation for particle 
masses as functions of $t=s^4$. Quite generally, 
renormalized equations for coefficients $c_t$ in 
$\cH_t$ may involve only masses, relative momenta, 
coupling constants, and the parameter $t$. Thus, 
in the elementary model, the equations involve only 
$\mu^2_t$, $\nu^2_t$, mass-mixing parameter $m^2_t$, 
and $t$ itself. These equations are independent of 
the particle momentum $p$. As a result, the RGPEP 
equations render a different representation of the 
same relativistic quantum theory for every value 
of $t$.

Each and every one of the effective theories
derived using the RGPEP, is defined in terms of a
different basis in the space of operators acting
in the Fock space. In the mass-mixing example, the
effective representations tend in the limit of $t
\rightarrow \infty$ to a relativistic theory of
free particles with masses $m_1$ and $m_2$. No
variation of the ground state $|0\rangle$ with $t$
is required in the procedure.

\subsection{ Standard, IF approach versus RGPEP }
\label{IFQ-RGPEP}

The comparison relies on a change of field variables 
in the classical Lagrangian of Eq.~(\ref{lagrangian}). 
The new variables are determined by diagonalization 
of the mass-squared matrix. The mass terms, 
\beq
- 2 \cL_{mass} \es 
  \mu^2 \phi^2  
+ \nu^2 \chi^2 
+ 2 m^2 \phi \, \chi
\eeq
can be written in the form of a $2 \times 2$
matrix sandwiched with a doublet of fields
$\Psi = [\phi, \chi]$. Namely, 
\beq
- 2\cL_{mass} 
\es 
\Psi^\dagger M^2 \Psi \\
\es     
\begin{array}{c}  \left[ \phi, \chi \right] \\ {} \end{array} 
\left[ \begin{array}{cc}    \mu^2  &    m^2 \\
                               m^2  &  \nu^2
               \end{array} \right] 
\left[ \begin{array}{c} \phi \\
                        \chi
               \end{array} \right] \, .
\eeq
The eigenvalues of matrix $M^2$, denoted 
by $m^2_1$ and $m_2^2$ above Eq.~(\ref{delta}) 
in Section \ref{analytic}, and the corresponding 
eigenvectors, are
\beq
m^2_{1,2}
\es (\mu^2 + \nu^2)/2 \pm
\sqrt{ (\mu^ 2 - \nu^2)^2/4  + m^4 } \, , \\
v_1 \es \left[ \begin{array}{r} \cos \varphi_\infty  \\
                              - \sin \varphi_\infty
               \end{array} \right] \, , 
\quad \quad 
v_2 \rs \left[ \begin{array}{r} \sin \varphi_\infty  \\
                              - \cos \varphi_\infty
               \end{array} \right] \, , 
\eeq 
where $\varphi_\infty$ is given in Eq.~(\ref{cinfty}).
Inverting the relation
\beq
\label{PsiDublet}
\Psi
\es
\xi \, v_1 + \zeta \, v_2 \, , 
\eeq
one can define the fields 
\beq
\label{xi1}
\xi   \es \cos \varphi_\infty \,\, \phi -  \sin \varphi_\infty \,\, \chi \, , \\
\label{zeta1}
\zeta \es \sin \varphi_\infty \,\, \phi +  \cos \varphi_\infty \,\, \chi  \, .
\eeq
This is a unitary change of field variables.
Since the terms that involve derivatives of
the fields $\phi$ and $\chi$ have equal 
coefficients in the classical Lagrangian 
density of Eq.~(\ref{lagrangian}), the density 
can be written as
\beq
\label{lagrangian1}
\cL \es {1 \over 2} \left[ (\partial \xi)^2 - m_1^2 \xi^2 \right] 
    +   {1 \over 2} \left[ (\partial \zeta)^2 - m_2^2 \zeta^2 \right] \, .
\eeq
This classical expression can now be quantized
from scratch in the IF of dynamics. 

The IF quantization involves definitions of the
fields $\xi$ and $\zeta$ and their conjugated
momenta. The quantization leads to a theory of
particles with masses $m_1$ and $m_2$ when one
defines the quantum fields $\xi$ and $\zeta$ and
their conjugated momenta $\pi_\xi$ and $\pi_\zeta$
using energy expressions $E_\xi^2(p) = m_1^2 + 
p^2$ and $E_\zeta^2 (p) = m_2^2 + p^2$ in defining 
the time derivatives of the fields, respectively. 
The new energy expressions guarantee that all terms 
of the type $a_{\xi p}^\dagger a_{\xi -p}^\dagger$ 
or $a_{\zeta p}^\dagger a_{\zeta - p}^\dagger$ 
cancel out in the Hamiltonian. Eqs.~(\ref{xi1}) 
and (\ref{zeta1}) imply that 
\beq
a_{\xi p} 
\es  \cos{\varphi_\infty} \, a_{0p} - \sin{\varphi_\infty} \, b_{0p} 
\label{axip} \, , \\
a_{\zeta p} 
\es  \sin{\varphi_\infty} \, a_{0p} + \cos{\varphi_\infty} \, b_{0p}  
\label{azetap} \, .
\eeq
These relations match Eqs.~(\ref{ainftyp}) and
(\ref{binftyp}). The matching shows that the IF
quantization of fields $\xi$ and $\zeta$ produces
the same result as the solution obtained entirely
in one quantum theory using the RGPEP, in which 
there is no need to re-quantize the theory due to 
inclusion of the mass-mixing interaction term. 

On the basis of knowing the full quantum
implications of the mass-mixing interaction term
in the FF of Hamiltonian dynamics, one can also 
write expressions for the initial IF quantum fields
$\hat \phi$, $\hat \chi$, $\hat \pi_\phi$, and
$\hat \pi_\chi$, in terms of the fields $\hat
\xi$, $\hat \zeta$, $\hat \pi_\xi$, and $\hat
\pi_\zeta$, using Eq.~(\ref{PsiDublet}) and right
energies for the time derivatives needed in $\hat
\pi_\xi$ and $\hat \pi_\zeta$. Substituting these
expressions into a classical IF Hamiltonian that
canonically corresponds to the Lagrangian density
of Eq.~(\ref{lagrangian}), one obtains the IF
quantum Hamiltonian that explicitly describes the
same physics as the FF quantum Hamiltonian
obtained from the RGPEP at $t=\infty$. The
vacuum-altering terms cancel out for all modes
with a finite momentum. 

However, when in a more complex theory than the
elementary example some additional interaction
terms cause divergences and other effects that are
difficult to see through, the IF quantization
approach may get stuck due to lack of a right
guess for the time derivatives. In contrast, the
RGPEP still indicates a direction for further
studies in realistic cases. Namely, while the free
theory that results from diagonalization of a
bilinear part in a Lagrangian density is certainly
not sufficient for establishing how to deal with
the IF vacuum problem, the RGPEP promises some
capability to work around the vacuum problem using
the FF.

The FF Hamiltonian at $t=s^4=0$, $\cP^-_0$,
involves fields $\phi$ and $\chi$. Their conjugate
``momenta,'' $\pi_\phi$ = $\partial^+ \phi$ and
$\pi_\chi$ = $\partial^+ \chi$, do not involve FF
time derivatives, i.e., they do not involve
derivatives with respect to $x^+$. Instead, the
``momenta'' are expressed through gradients of 
the fields in the front hyper-plane. Rotation of 
the quantum fields automatically rotates the 
quantum ``momenta.'' 

By the way, the fields $\xi$ and $\zeta$ can be
used as initial variables also in the FF. The
RGPEP provides no additional value in such setup,
since there is no interaction between the free
fields $\xi$ and $\zeta$. However, when more
interactions are added, nothing prevents the RGPEP
from application to the whole quantum theory using
the effective particle operators associated with
the fields $\xi$ and $\zeta$, instead of $\phi$ and
$\chi$.

It should be mentioned that an interesting example
of the IF application of a similarity
renormalization group procedure in a fixed source
model has been recently considered by Jones and
Perry~\cite{JonesPerry}. In the fixed source
model, the interaction term is only linear in the
quantum field variables, different momentum modes
evolve separately, and one obtains the well-known
solution in an elegant way. The fixed source model
does not appear to suggest how to proceed in the
IF when interaction terms involve more than one
field and create a genuine vacuum problem.

\section{ Conclusion }
\label{conclusion}

The case of a theory with two free fields with a
mass-mixing interaction term can be generalized 
to theories with an arbitrary number $n > 2$ of
fields and mass mixing terms. In such theories,
the RGPEP equation describes the evolution of a 
mass matrix of dimension $n \times n$ with $s$. 
The solution tends for $s \rightarrow \infty$ to 
a diagonal matrix, whose eigenvalues provide 
physical masses for $n$ species of free particles. 

Degeneracy of the mass matrix, which may
correspond to a symmetry in a theory, prevents its
full diagonalization via the RGPEP equation. In
this case, an artificial infinitesimal breaking of
the degeneracy can be introduced in order to
enable RGPEP to identify a solution in the limit
$s \rightarrow \infty$, as the artificial breaking
is being removed. 

The diagonalization of the mass matrix does not
correspond to a minimization of a classical
potential in the IF. Instead, it corresponds to
identification of the eigenmodes in classical
field oscillations. One has to use the eigenmodes
in the IF quantization procedure in order to solve
a vacuum problem in the absence of interactions
other than the mass mixing. However, when such
additional interaction terms, involving products
of more than two fields, are included in a theory,
the IF vacuum problem can no longer be solved
using the field combinations that correspond to
eigenvectors of the mass matrix. The additional
interactions typically contribute to particle
masses, bound states may develop, and, as it would
have to happen in the case of confinement, the
full theory eigenmodes do not even correspond to
the fields present in an initial Lagrangian. 

The intriguing feature of the RGPEP, illustrated
here in the elementary model with a mass-mixing
interaction term, is that it applies to quantum
theories via steps that are essentially
independent of the type of interaction one
grapples with, while the vacuum problem is treated
in a new way. Namely, the vacuum stays simple
while the interaction terms evolve towards 
expressions in terms of effective degrees of
freedom. This feature makes the RGPEP a deserving
candidate for application to more realistic
theories than the elementary model discussed here.
It is evident from the works referenced in this
article that the RGPEP can be applied to realistic
quantum field theories. The elementary example
described here is thus of interest not only as an
illustration of an exact non-perturbative solution
of the RGPEP equations but also as the indicator of 
a difference between the options one has got left 
for treating vacuum problems in the IF and FF of 
Hamiltonian dynamics in relativistic quantum field 
theories.

\begin{appendix}

\section{ Solution for $\cH_f$ dependent on $t$ }
\label{massess}

Discussion of Eq.~(\ref{masssquared}) in
Section \ref{boostinvariance} included the
case of $\cH_f$ containing masses dependent
on $t$, which yields a $2 \times 2$ mass-squared 
matrix equation of the form
\beq
\left[ \begin{array}{cc}
               \mu_t^2  &    m_t^2      \\
                 m_t^2  &  \nu_t^2
               \end{array} \right]'
\es
\left[
\left[  \left[ \begin{array}{cc}
               \mu_t^2  &  0            \\
               0      &  \nu_t^2
               \end{array} \right],
\left[ \begin{array}{cc}
               \mu_t^2  &   m_t^2       \\
               m_t^2    &   \nu_t^2
               \end{array} \right] \right],
\left[ \begin{array}{cc}
               \mu_t^2 &  m_t^2      \\
                m_t^2  &  \nu_t^2
               \end{array} \right]
\right] \, .
\eeq
This equation matches the Wegner equation for 
a Hamiltonian matrix~\cite{Wegner1} of a 
two-level system. Its analytic solution is 
well-known but as far as the author knows it 
was never considered before in the context 
of particle masses in an exactly soluble 
quantum field theory in the FF of dynamics.

Proceeding as in Section \ref{analytic}, one 
obtains
\beq
\label{deltaW}
\delta'   
\es   4 \, \delta \, \gamma^2 \, , \\
\label{gamma1W}
\gamma'
\es  - \delta^2 \, \gamma \, .
\eeq
Multiplying the first of these two equations 
by $2\delta$ and the second by $2\gamma$, one 
arrives at 
\beq
\label{delta2W}
{\delta^2}'   
\es   8 \, \delta^2 \, \gamma^2 \, , \\
\label{gamma2W}
{\gamma^2}'
\es  - 2 \delta^2 \, \gamma^2 \, ,
\eeq
which implies the same constant 
$\epsilon^2 = \delta^2 + 4 \gamma^2$
as in Section \ref{analytic}.
After eliminating $\gamma^2$ from Eq.~(\ref{delta2W}),
\beq
\label{delta3W}
{\delta^2}' \es  2 \,  \delta^2  \, (\epsilon^2 - \delta^2) \, .
\eeq
The solutions corresponding to Eqs.~(\ref{deltamut}) 
and (\ref{mt}), are 
\beq
\delta \mu^2_t \es 
\delta \mu^2 \, {\epsilon \, e^{x_t} \over
                 \sqrt{ \epsilon^2 -1 + e^{2 x_t} } } \, , \\
m_t^2          \es 
         m^2 \, {\epsilon          \over
                 \sqrt{ \epsilon^2 -1 + e^{2 x_t} } } \, ,
\eeq
where $x_t= (\delta m^2)^2 \, t$. 

Since the generator given in Eqs.~(\ref{GP}) 
and (\ref{GP1}) is now altered to contain the
varying $\delta \mu_t^2$ instead of the constant 
$\delta \mu^2$, the angle $\varphi_t$ given by Eq. 
(\ref{ctsolution}) is replaced by 
\beq
\label{ctsolutionW1}
\varphi_t 
\es 
- \int_0^t \delta \mu^2_\tau \, m_\tau ^2 \, d\tau  
\label{ctsolutionW2}
\rs
{1 \over 2} \,  \left( \arctan{ 1 \over \sqrt{ \epsilon^2 - 1} }
                     - \arctan{ e^{ (\delta m^2)^2 t} \over \sqrt{ \epsilon^2 - 1} } \right) \, .
\eeq 
This result deviates from the result in Eq.
(\ref{ctsolution2}) for finite values of $t$. The
difference in the angles of rotation, $\varphi_t$,
implies different combinations of operators
$a_{0p}$ and $b_{0p}$ in Eqs.~(\ref{atp}) and
(\ref{btp}) for the same $t$. This means that the
effective particle operators at any finite $t > 0$
depend on the choice of the generator, although
the Hamiltonians as operators are just one and the
same operator for all values of $t$ and both
choices of the generator. When $t \rightarrow
\infty$, Eqs.~(\ref{ctsolution}) and
(\ref{ctsolutionW1}) produce the same result for
$\varphi_\infty$ for arbitrary values of $\epsilon > 1$,
\beq
\label{cinftyA}
\varphi_\infty 
\es 
               \arctan{ \sqrt{ \epsilon + 1 \over \epsilon - 1} } 
             - \pi/2 
\rs
{1 \over 2} \, \arctan{ 1 \over \sqrt{ \epsilon^2 - 1} }
             - \pi/4 \, .
\eeq
Thus, the change in the generator from a 
constant $\cH_f$ to a $t$-dependent full 
free part of $\cH_t$, does not lead to any
change in the effective particles that one 
obtains for $t \rightarrow \infty$ as a 
solution of the theory.

\end{appendix}



\end{document}